# Investigation of Multifractal Properties of Additive Data Stream

Igor Ivanisenko [1], Lyudmyla Kirichenko [2], Tamara Radivilova [3]

[1] Kharkiv National University of Radioelectronics, Nauky Avenue 14, Kharkiv, 61166, UKRAINE,
e-mail: ihor.ivanisenko@nure.ua

[2] Kharkiv National University of Radioelectronics, Nauky Avenue 14, Kharkiv, 61166, UKRAINE
e-mail: lyudmyla.kirichenko@nure.ua

[3] Kharkiv National University of Radioelectronics, Nauky Avenue 14, Kharkiv, 61166, UKRAINE
e-mail: tamara.radivilova@nure.ua

*Abstract* – **The work presents results of a numerical study of fractal characteristics of multifractal stream at addition of stream, which does not have multifractal properties. They showed that the generalized Hurst exponent of total stream tends to one of original multifractal stream with increase in signal/noise ratio.**

## I. INTRODUCTION

Numerous researches of processes in a network have shown that statistical characteristics of the traffic have property of time scale invariance (self-similarity). Self-similar properties were discovered in the local and global networks, particularly traffic Ethernet, ATM, applications TCP, IP, VoIP and video streams. The reason for this effect lies in the features of the distribution of files on servers, their sizes, and the typical behavior of users. It was found that initially not having self-similarity data streams passing on nodal processing servers and active network elements became self-similar.

The self-similar traffic has the special structure that preserves on many scales. There are always a number of extremely large bursts at relatively small average level of the traffic. These bursts are cause significant delays and losses of packages, even when the total load of all streams are more less than maximal values. In a classical case for Poisson stream buffers of an average size will be enough. The queue can be formed in short-term prospect, but for the long period buffers will be cleared. However in a case of self-similar traffic queues have more greater length.

Now the multifractal properties of traffic are intensively studied. Multifractal traffic is defined as an extension of self-similar traffic due to take account of properties of second and higher statistics.

Thus, an important task to improve the network quality of service is the study of self-similar and multifractal properties of data streams. A characteristic feature of computer networks is the multiplexing of streams, so characteristics of the additive self-similar streams have special significance. In [1-2], theoretical and numerical properties of self-similar additive processes were studied. It was shown that sum of several self-similar processes with different values of the Hurst exponent has maximal one. In [3-6], the results of experimental studies of the properties of additive data traffic which confirm the theoretical results are presented. However, these studies do not take into account multifractal properties of streams, the quantitative characteristic of which is the generalized Hurst exponent.

The purpose of the present work is to research numerically the changing in fractal characteristics of multifractal stream at addition of stream, which does not have multifractal properties.

## II. CHARACTERISTICS OF SELF-SIMILAR AND MULTIFRACTAL PROCESSES

Stochastic process $X(t)$, $t \geq 0$ with continuous real-time variable is said to be self-similar of index $H$, $0 < H < 1$, if for any value $a > 0$ processes $X(at)$ and $a^{-H}X(at)$ have same finite-dimensional distributions:

$$\text{Law}\{X(at)\} = \text{Law}\{a^H X(t)\}. \qquad (1)$$

The notation $\text{Law}\{\cdot\}$ means finite distribution laws of the random process. Index $H$ is called Hurst exponent. It is a measure of self-similarity or a measure of long-range dependence of process. For values $0{,}5 < H < 1$ time series demonstrates persistent behaviour. In other words, if the time series increases (decreases) in a prior period of time, then this trend will be continued for the same time in future. The value $H = 0.5$ indicates the independence (the absence of any memory about the past) time series values. The interval $0 < H < 0.5$ corresponds to antipersistent time series: if a system demonstrates growth in a prior period of time, then it is likely to fall in the next period.

The moments of the self-similar random process can be expressed as

$$E\left[|X(t)|^q\right] = C(q) \cdot t^{qH} \quad (2)$$

where the quantity $C(q) = E\left[|X(1)|^q\right]$.

In contrast to the self-similar processes (1) multifractal processes have more complex scaling behavior:

$$\text{Law}\{X(at)\} = \text{Law}\{\mathcal{M}(a) \cdot X(t)\} \quad (3)$$

where $\mathcal{M}(a)$ is random function that independent of $X(t)$. In case of self-similar process $\mathcal{M}(a) = a^H$.

For multifractal processes the following relation holds:

$$E\left[|X(t)|^q\right] = c(q) \cdot t^{qh(q)} \quad (4)$$

where $c(q)$ is some deterministic function, $h(q)$ is generalized Hurst exponent, which is generally non-linear function. Value $h(q)$ at $q = 2$ is the same degree of self-similarity $H$. Generalized Hurst exponent of monofractal process does not depend on the parameter $q$: $h(q) = H$.

Fig. 1 shows plot example of generalized Hurst exponent $h(q)$ for monofractal and multifractal stochastic processes. In the case of a monofractal process Hurst exponent is a straight line.

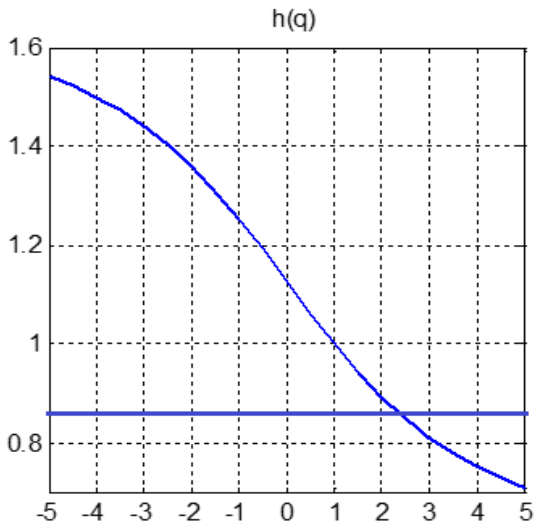

Figure 1. Generalized Hurst exponent for monofractal (straight) and multifractal (curve) stochastic processes.

## III. MODELS OF SELF-SIMILAR AND MULTIFRACTAL DATA TRAFFIC

The main tool for the study and predict the behavior of self-similar data streams is simulation, which requires a model of self-similar and multifractal input load.

Fractional Brownian motion (fBm) with a parameter $H$ is often considered as a stochastic process possessing self-similar properties and it is widely used in the theory of network traffic [7]. fBm with a parameter $H = 0.5$ is ordinary Brownian motion. The increment process of fBm is known as fractional Gaussian noise (fGn). Theoretically fGn can be considered as a model of self-similar traffic with a defined Hurst exponent and the corresponding long-term dependence. However, this model has a number of shortcomings one of which is zero mean and negative values.

Easy transition from fGn to self-similar traffic with positive values is the transformation that has been proposed in [8]. In the simplest case, the modeling traffic realization is

$$Y(t) = \text{Exp}[fGn(t)]. \quad (5)$$

The proposed transformation preserves the long-term dependence of the stochastic process. The stochastic process $Y(t)$ is a self-similar stochastic process as the same Hurst exponent $H$, as the initial fractal Gaussian noise. The variable $Y(t)$ has a log-normal distribution.

A suitable model of the traffic with predetermined multifractal properties are stochastic cascade processes. It was proposed to use for modeling multifractal traffic realizations of the stochastic multiplicative binomial cascade [7, 9]. In the construction of stochastic cascades the weight coefficients are independent values of a random variable. In [9] the beta-distribution random variable was used for weights. This allows to generate the trafic realizations with varying degree of heterogeneity, i.e. with a large range of multifractal properties.

Graph of typical multifractal cascade realization is represented at the top of the Fig. 2. In the middle part of the Fig. 2 the realization of traffic, which has independent values ($H = 0.5$) and generates according to (5) is shown. At the bottom of the Fig. 2 the realization of the total stream is shown.

Modern information networks are built on the multiplex data streams. The mechanism of statistical multiplexing of information streams is widely used in telecommunications, because it allows to economical use of the bandwidth of the main channels. It consists in the fact that the individual sources are added streams in the main channel with saving bandwidth.

We wondered how the multifractal characteristics of the total stream change. In each case the basic stream were multifractal cascade realizations, to which the realizations of various probability and correlation properties were added. They were exponents of white noise obtained in accordance with (5), exponent of fGn having a long-term dependence, autoregressive time series having short term memory and series of various distribution independent random values.

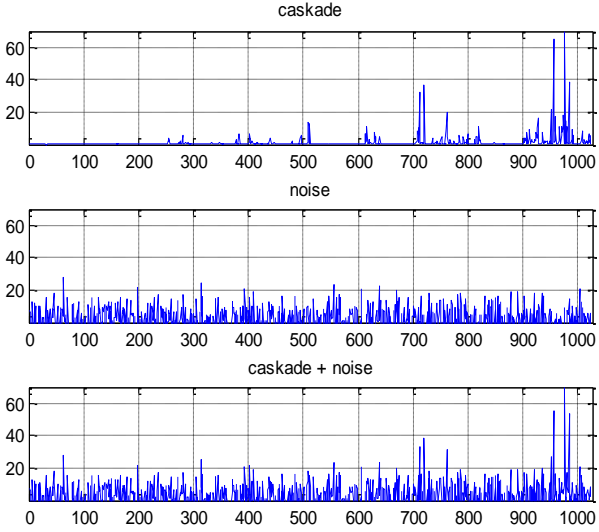

Figure 2. Model realizations: multifractal, exponents of white noise and total

## IV. RESEARCH RESULTS

In the work the investigation of multifractal properties of total streams of various types was carried out.

The investigated model additive stream is introduced as

$$X_{SUM}(t) = X_{MULTI}(t) + X_{NOISE}(t), \qquad (6)$$

where $X_{MULTI}(t)$ is realization of multifractal cascade process, $X_{NOISE}(t)$ is additive noise realization. The factor $SNR$ is the factor which characterizes the ratio of multifractal stream and noise one:

$$SNR = \text{Var}[X_{MULTI}] / \text{Var}[X_{NOISE}]. \qquad (7)$$

Consider how the generalized Hurst exponent $h_{SUM}(q)$ of total stream changes in the case where the additive $X_{NOISE}(t)$ is the exponent of white noise. Fig. 3 shows the generalized Hurst exponent $h(q)$ in the range of parameter $-10 \le q \le 10$. The top solid line corresponds to $h_{MULTI}(q)$ of multifractal realization $X_{MULTI}(t)$; dotted line 1 corresponds to $h_{SUM}(q)$ for total realization $X_{SUM}(t)$ when value $SNR = 5$ and dotted line 2 is $h_{SUM}(q)$ in case $SNR = 1$. It is obvious that the generalized Hurst exponent $h_{SUM}(q)$ and the Hurst exponent $h_{MULTI}(q)$ of the original multifractal stream $X_{MULTI}(t)$ in case 1 are very close for positive values of the parameter $q$. This means that the multifractal properties of original stream do not change and can be easily identified.

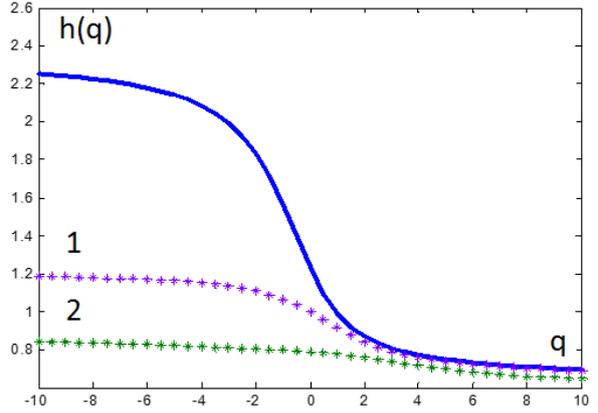

Figure 3. Generalized Hurst exponents of the original multifractal stream and total streams with different signal/noise ratio.

Taking into account these results, further analysis of the generalized Hurst exponent $h(q)$ only the positive values of the parameter $q$ were considered. Numerical research showed that at signal/noise ratio $SNR \ge 5$ the generalized Hurst exponent $h_{SUM}(q)$ of total stream and $h_{MULTI}(q)$ of the original multifractal stream are almost identical at values $q \ge 0$.

Generalized Hurst exponent $h_{SUM}(q)$ was numerically investigated by changing the signal / noise ratio $SNR$. It is shown that with decreasing the ratio value $SNR$ of 5 to 1 generalized Hurst exponent $h_{SUM}(q)$ tends to the $h_{MULTI}(q)$ of original stream. Fig. 4 shows the function $h_{MULTI}(q)$ of the original multifractal realization (●-line) and $h_{SUM}(q)$ of total ones when a number of the coefficient $SNR = 2$ (*-line), 4 (∆-line), 5 (š-line) and 10 (¾-line).

Numerical studies have shown that this relation holds for all kinds of additive traffic: exponent of fGn with different Hurst parameter ($H > 0.5$), traffic realization having autoregressive dependence. The case when the additive traffic does not possess the self-similar properties and does not have the normal distribution was also investigated.

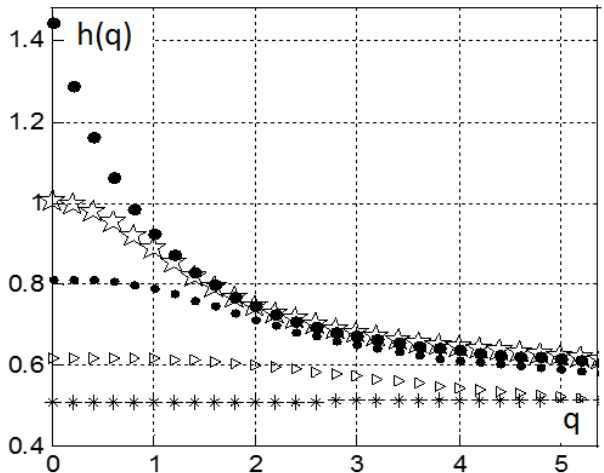

Figure 4. . Generalized Hurst exponents of the original multifractal stream (●-line) and total ones with different signal/noise ratio: 2 (∗-line), 4 (Δ-line), 5 (Š-line) and 10 (¾-line) in case of self-similar noise.

Fig. 5 shows the function $h_{MULTI}(q)$ of the original multifractal realization (●-line) and $h_{SUM}(q)$ of total ones with different signal/noise ratio when the additive traffic is independent values of uniform distribution random variable. In this case the lines of $h_{SUM}(q)$ correspondent at $SNR = 2$ (Δ-line), 4 (š-line) and 5 (¾-line).

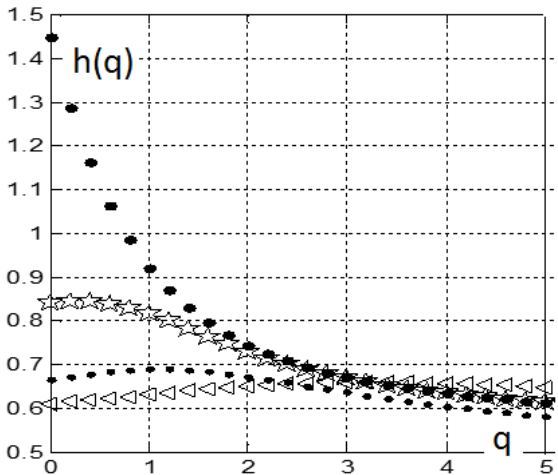

Figure 5. Generalized Hurst exponents of the original multifractal stream (●-line) and total ones with different signal/noise ratio: 2 (Δ-line), 4 (Š-line) and 5 (¾-line) in case of not self-similar noise.

Obviously, in the case of uncorrelated not self-similar additive stream $X_{NOISE}(t)$ the total one $X_{SUM}(t)$ has the multifractal properties of the original stream $X_{MULTI}(t)$ at smaller values of the ratio $SNR$.

## V. CONCLUSION

The work presents results of a numerical study of the changing in fractal characteristics of multifractal stream at addition of stream, which does not have multifractal properties. The study results showed that the fractal characteristics of multifractal stream are saved depending on the magnitude of signal/noise ratio. With increase in signal/noise ratio the generalized Hurst exponent of total stream tends to one of original multifractal stream in the region of positive values of the parameter. If additive stream does not have self-similar properties, multifractal characteristics hold at smaller ratio.

Further work is expected to investigate the other models of self-similar processes and to study characteristics of the total multifractal data streams. The research results have applied significance not only in telecommunications but also in radio engineering and digital seismology, where one of the main problems is of the useful signal in a noisy environment.